\documentstyle[10pt,aas2pp4,psfig]{article}

\received{}
\revised{}
\accepted{}

\journalid{}{}
\articleid{}{}
\paperid{}

\newcommand{\Ia}{\mbox{I{\tiny A}}}
\newcommand{\Ib}{\mbox{I{\tiny B}}}
\newcommand{\Ic}{\mbox{I{\tiny C}}}
\newcommand{\Id}{\mbox{I{\tiny D}}}
\newcommand{\Ie}{\mbox{I{\tiny E}}}
\newcommand{\If}{\mbox{I{\tiny F}}}
\newcommand{\Ide}{\mbox{I{\tiny D/E}}}
\newcommand{\Ta}{\mbox{T$_1${\tiny A}}}
\newcommand{\Tb}{\mbox{T$_1${\tiny B}}}
\newcommand{\Tc}{\mbox{T$_1${\tiny C}}}
\newcommand{\Tbc}{\mbox{T$_1${\tiny B/C}}}

\lefthead{Perrett et al.}
\righthead{The GCSs of NGC 1400 \& NGC 1407}

\begin{document}

\small{

\title{The Globular Cluster Systems of NGC~1400 and NGC~1407}

\author{Kathryn M. Perrett, David A. Hanes\altaffilmark{1,2},
Steven T. Butterworth\altaffilmark{1} and JJ Kavelaars\altaffilmark{2}}
\affil{Department of Physics, Queen's University, Kingston, Ontario
K7L 3N6, Canada \\ 
E-mail: perrett@astro.queensu.ca, hanes@astro.queensu.ca,
bworth@astro.queensu.ca, kauelaar@astro.queensu.ca}

\altaffiltext{1}{Visiting Astronomer, Cerro Tololo Inter-American Observatory,
National Optical Astronomy Observatories, operated by the Association
of Universities for Research in Astronomy (AURA), Inc., under contract
with the National Science Foundation (NSF).}

\altaffiltext{2}{Visiting Astronomer, Canada-France-Hawaii Telescope, operated
by the National Research Council of Canada, le Centre National de la Recherche
Scientifique de France, and the University of Hawaii.}

\author{Doug Geisler}
\affil{National Optical Astronomy Observatories, Kitt Peak National
Observatory \\ E-mail:  doug@noao.edu}

\vskip 0.5cm
\and

\author{William E. Harris}
\affil{Department of Physics and Astronomy, McMaster University, Hamilton,
Ontario L8S 4M1, Canada \\ 
E-mail:  harris@azalea.physics.mcmaster.ca}

\vskip 0.5cm

\begin{abstract}
{The two brightest elliptical galaxies in the Eridanus~A group, NGC
1400 and NGC 1407, have been observed in both the Washington $T_1$ and
Kron-Cousins $I$ filters to obtain photometry of their globular
cluster systems (GCSs).  This group of galaxies is of particular
interest due to its exceptionally high $M/L$ value, previously
estimated at $\sim 3000h$, making this cluster highly
dark-matter-dominated (Gould 1993).  NGC 1400's radial velocity
(549~km/s) is extremely low compared to that of the central galaxy of
Eridanus A (NGC 1407 with $v_\odot=1766$~km/s) and the other members
of the system, suggesting that it is a foreground galaxy projected by
chance onto the cluster.  Using the shapes of the globular cluster
luminosity functions, however, we derive distances of $17.6 \pm
3.1$~Mpc to NGC~1407 and $25.4 \pm 7.0$~Mpc to NGC~1400.  These
results support earlier conclusions that NGC 1400 is at the distance
of Eridanus~A and therefore has a large peculiar velocity.  Specific
frequencies are also derived for these galaxies, yielding values of
$S_N=4.0 \pm 1.3$ for NGC 1407 and $S_N = 5.2 \pm 2.0$ for NGC 1400.
In this and other respects, these two galaxies have GCSs which are
consistent with those observed in other galaxies.}
\end{abstract}

\keywords{Globular cluster systems, galaxy clusters, dark matter}
\vfill\eject

\section{Introduction}

NGC 1400, a bright E0/S0 galaxy located in a large sub-condensation of
the Eridanus cluster, has recently drawn attention to itself and its
curious environment.  The southern Eridanus group of galaxies extends
between $3^h15^m \lesssim$ R.A. $\lesssim 4^h$ and $-26\arcdeg
\lesssim$ Dec.  $\lesssim -15\arcdeg$, and exhibits a large degree of
internal sub-clustering (see Figure~1 from Willmer et al. 1989).  Its
most concentrated clump of galaxies is Eridanus A ($3^h40^m,
-19\arcdeg$), which includes roughly 50 galaxies within a radius of
$\sim0\fdg8$.

At an estimated distance of 16.4~Mpc (Tonry 1991), the large central
E0 galaxy of the Eridanus~A sub-cluster (NGC 1407) has a heliocentric
velocity typical of the majority of the members of the group.  NGC~1400,
however, reveals an anomalously low redshift (549 km/s)
compared with those of its neighbouring galaxies, causing some uncertainty
as to its true distance.
Table~\ref{tab1} lists the members of Eridanus~A for which
heliocentric velocities have been determined.  The mean
radial velocity of these galaxies is $1643 \pm 95$ km/s, or $1765 \pm 93$ km/s
if NGC~1400 is not considered a member.

\begin{deluxetable}{lccrc}
\tablewidth{275pt}
\tablecaption{\small Galaxies of Eridanus~A with known velocities
(Willmer et al. 1989).  \label{tab1}}
\tablecolumns{5}
\tablehead{
\colhead{Galaxy} & \colhead{R.A.} & \colhead{Dec.} & \colhead{$v_\odot$} &
\colhead{B} \nl
\colhead{~} & \colhead{(1950)} & \colhead{(1950)} &
\colhead{(km/s)} & \colhead{~}
}
\startdata
NGC~1383 & 3 35 23 & -18 30.1 & 1948$\pm$19 & 14.0 \nl
NGC~1390 & 3 35 37 & -19 10.3 & 1215$\pm$33 & 14.6 \nl
NGC~1393 & 3 36 23 & -18 35.4 & 2185$\pm$26 & 13.9 \nl
NGC 1400\tablenotemark{a} & 3 37 16 & -18 51.0 & 549$\pm$21 & 12.1 \nl
IC~343 & 3 37 52 & -18 36.3 & 1869$\pm$30 & 14.5 \nl
NGC 1407\tablenotemark{b} & 3 37 57 & -18 44.5 & 1766$\pm$21 & 11.1 \nl
ESO~548-G48 & 3 38 04 & -19 05.5 & 1817$\pm$27 & 14.4 \nl
IC~346 & 3 39 29 & -18 25.6 & 1897$\pm$47 & 14.1 \nl
ESO~548-G79 & 3 39 41 & -19 03.2 & 2053$\pm$31 & 14.4 \nl
ESO~549-G02 & 3 40 43 & -19 10.8 & 1134$\pm$36 & 15.0 \nl
\enddata
\tablenotetext{a,b}{~The galaxies in this study}
\end{deluxetable}

This straightforward question as to the true distance of a potential
Eridanus~A member has resulted in what may prove to be an extraordinary
find:  the sub-cluster has an abnormally high $M/L$ value.
Of the 10 Eridanus~A galaxies listed in Table~\ref{tab1}, the two
early-type galaxies NGC 1407 and NGC 1400 account for nearly 80\% of the
light.  Adding the blue luminosities of the Eridanus galaxies within $0\fdg8$
of NGC 1407 (including NGC 1400) and calculating the mean of four virial
mass estimators, Gould (1993) finds $M/L \sim 3125 h$.

This value is anomalously large; typically,
groups or clusters of galaxies have $M/L$ values of several hundred,
making Eridanus~A one of the darkest clusters known.
Considering only the virial theorem mass, the exclusion of NGC 1400
from the cluster mass calculation reduces $M/L$ by roughly a factor of 2.
Consequently, even if NGC 1400 has not virialized within the cluster,
Eridanus~A's $M/L$ is still excessively high.

If NGC 1400 is at the distance of Eridanus~A, it may yet have
originated from some other location, giving rise to a peculiar motion
which is not due to the mass concentration of the sub-cluster.  Since
there are no other nearby mass concentrations large enough to account
for NGC 1400's motion, this possibility is unlikely (see Gould 1993).
Therefore, if NGC 1400 is at the distance of Eridanus~A, it is likely
to be dynamically associated with the sub-cluster.  Is NGC~1400 indeed
at the distance of Eridanus~A, or is it instead a foreground galaxy
mistakenly associated with the group due to projection effects?

Five independent studies have attempted to determine whether NGC 1400
is at the distance of Eridanus~A.  The results of four of these
studies are summarized in Table~\ref{tab2} (see discussion in Gould 1993 and
references therein).  One additional piece of evidence presented by
Gould is the globular cluster luminosity function (GCLF), credited as
showing that NGC 1400 and NGC 1407 are at the same distance.  However,
these GCLFs are secondary results obtained by Tonry (1991) during the
surface-brightness-fluctuation (SBF) analysis, and no rigorous
investigation of the globular cluster systems of the two bright
galaxies in Eridanus~A has yet been performed.

\begin{deluxetable}{cccl}
\tablewidth{0pt}
\tablecolumns{4}
\tablecaption{\small Distance Measurements of NGC~1400 and NGC~1407 (Gould 1993).
\label{tab2}}
\tablehead{
\colhead{Method} & \colhead{NGC~1400} & \colhead{NGC~1407} &
\colhead{Reference} \nl
\colhead{~} & \colhead{Distance (Mpc)} & \colhead{Distance (Mpc)} &
\colhead{~}
}
\startdata
SBF & $16.3 \pm 1.0$ & $16.4 \pm 1.0$ & Tonry 1991 \nl
$D_n-\sigma$ & $(24 \pm 1)h^{-1}$ & $(19\pm 1)h^{-1}$ & Faber et al. 1989 \nl
X-ray/$B$ & $(16 \pm 2)h^{-1}$ & $(18 \pm 1)h^{-1}$ & Donnelly et al. 1990 \nl
$(u-V)/V$ & $(40 \pm 2)h^{-1}$ & $(30 \pm 2)h^{-1}$ &
Sandage \& Visvanathan 1978; \nl
~ & ~ & ~ & Visvanathan \& Sandage 1977 \nl
\enddata
\end{deluxetable}

We therefore elect to use the globular cluster luminosity functions of
the two brightest ellipticals in Eridanus~A, NGC 1400 and the central
galaxy NGC 1407, to provide new estimates of their distances.
In \S\ref{sec:obs} and \S\ref{sec:reduc} we discuss the observations
and data reduction, while in \S\ref{sec:gclf} we present the GCLFs and
distance determinations.  In addition, the inferred globular
cluster radial profiles and scaled population
sizes of these two galaxies are investigated in \S\ref{sec:rp} and
\S\ref{sec:sp} respectively, in search of any
signature of the peculiar environment in which they are found.

\section{Observations}
\label{sec:obs}

The data for this study of the globular cluster systems (GCSs) of NGC
1400 and NGC 1407 were acquired as part of two separate
observing runs.  The first set of observations was obtained on the
nights of November 14 and 16, 1993, with the Cerro Tololo
Inter-American Observatory (CTIO) 4-m telescope.  Use of the Tek
$2048\times2048$ CCD camera at prime focus with an image scale of
$0\farcs46$ per pixel yielded an image size of
$15\farcm7\times15\farcm7$.  The readout noise was 6 electrons/pixel,
and the gain was set to 3.2 electrons per ADU.  A total of three
images, each one including both NGC 1400 and NGC 1407, were taken
using the Washington $T_1$ filter (Canterna 1976; Harris \& Canterna
1977).  The seeing was mediocre at best, ranging from $1\farcs5$ to
$2\farcs0$ over the two nights.

The second set of data was collected on January 2 and 3, 1995, with
the Canada-France-Hawaii 3.6-m telescope (CFHT).  The Loral3 detector
($2048\times2048$) was used at a nominal gain of 1.45 electrons/ADU
and with a readout noise of 9 electrons/pixel.  This detector, mounted
behind the re-imaging optics of the MOS instrument, yielded an image
scale of $0\farcs32$ pix and a frame size of
$10\farcm9\times10\farcm9$.  Six images, three each of NGC 1407 and
NGC 1400, were obtained using the Kron-Cousins $I$-band filter. The
seeing varied from $1\farcs1$ to $1\farcs9$ over the two nights.  A
log of the observations is presented in Table~\ref{tab3}.  In addition
to the program fields, standard star fields, dome flatfields and bias
frames were also obtained.

\begin{deluxetable}{cccccccc}
\tablecaption{\small Record of Observations. \label{tab3}}
\tablewidth{0pt}
\tablecolumns{8}
\tablehead{
\colhead{Frame} & \colhead{Object(s)} & \colhead{Date} & \colhead{Start UT} &
\colhead{Exp. (sec)} & \colhead{Filter} & \colhead{Airmass} & \colhead {Seeing
($\arcsec$)}
}
\startdata
\Ta & N1400/1407 & Nov.~14/93 & 08:21 & 600 & $T_1$ & 1.534 & 1.5 \nl
\Tb & N1400/1407 & Nov.~16/93 & 07:28 & 900 & $T_1$ & 1.294 & 1.9 \nl
\Tc & N1400/1407 & Nov.~16/93 & 07:50 & 900 & $T_1$ & 1.394 & 2.0 \nl
\Ia & N1407 & Jan.~2/95 & 08:48 & 600 & $I$ & 1.447 & 1.4 \nl
\Ib & N1407 & Jan.~2/95 & 09:01 & 600 & $I$ & 1.289 & 1.9 \nl
\Ic & N1407 & Jan.~3/95 & 08:01 & 900 & $I$ & 1.345 & 1.1 \nl
\Id & N1400 & Jan.~2/95 & 08:20 & 600 & $I$ & 1.367 & 1.5 \nl
\Ie & N1400 & Jan.~2/95 & 08:33 & 600 & $I$ & 1.397 & 1.5 \nl
\If & N1400 & Jan.~3/95 & 07:37 & 900 & $I$ & 1.290 & 1.2 \nl
\enddata
\end{deluxetable}

\section{Data Reduction}
\label{sec:reduc}


Preprocessing of the raw frames included bias subtraction and division
by a mean dome flatfield exposure.  Frames taken in comparable seeing
were combined to form composite images with improved signal-to-noise.
The resultant composite images with substantially different seeing
were reduced individually using the standard Sun/Unix implementation of
IRAF\footnotemark\footnotetext{Image Reduction and Analysis Facility,
distributed by the National Optical Astronomical Observatories, which
is operated by AURA under contract with the NSF.}.  Figure~\ref{fig1}
shows an example of a preprocessed frame in the $T_1$ filter.

\begin{figure}
\hspace*{\fill}\psfig{file=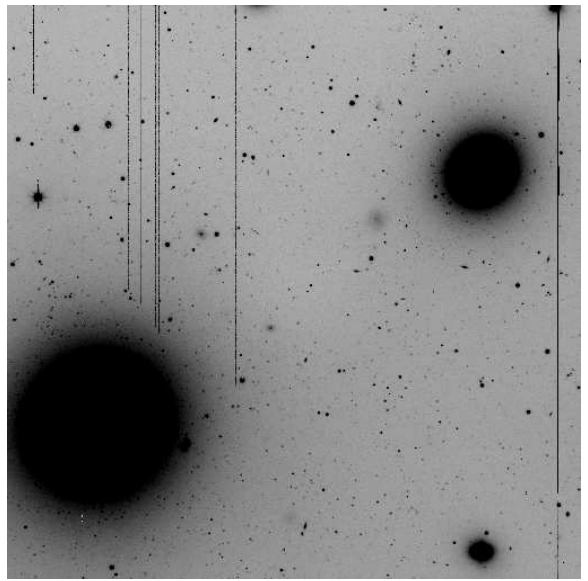,height=3in,width=3in}
\hspace*{\fill}\caption{\small A preprocessed image of NGC 1400
(upper right) and NGC 1407 (lower left) in the $T_1$ filter from CTIO.
The size of the field is $15\farcm7\times15\farcm7$ with north at the
bottom and east at the left of the frame.}
\label{fig1}
\end{figure}


The data reduction was largely automated with the aid of an IRAF script
and the DAOPHOT subroutines (Stetson 1987).
A median filtering procedure was applied in order to create a model of
the diffuse galaxy light which was then subtracted from the original
image to better reveal the underlying globular cluster system.
The fitting of a model stellar profile (a ``point spread function'' -- PSF)
was carried out upon the filtered image to yield the uncalibrated
photometry list.


The fraction of objects successfully detected in each frame as a function of
magnitude is known as the completeness function (Harris 1990).
Using the IRAF ADDSTAR task, a total of $10^3$ artificial stars of known
magnitudes were added to each frame in 10 trials, thus increasing the
population of objects in an individual frame by roughly 20\% per trial.
In order to compensate for the variation in object
detection with radial distance
from the inner galaxy regions, separate completeness functions were
determined for a series of annuli centred on the galaxy.

The uncertainties associated with the computed completeness function ($f$)
are derived assuming a binomial distribution:

\begin{equation}
\sigma^2_f\approx \frac{f(1-f)}{n_{\rm add}},
\label{eq:cfunc}
\end{equation}

\noindent where $n_{\rm add}$ refers to the number of artificial stars added
to a given bin (Bolte 1989).
The detection limits ($50\%$ recovery levels) for the program frames
are listed in Table~\ref{tab4}.
Using the known magnitudes of the artificial stars, we found that 
the systematic errors in the recovery process are negligible.  
An estimate of the internal uncertainty
as derived from the scatter of the artificial stars
was within $0.01\lesssim\sigma\lesssim0.05$ for all frames over the
magnitude range of interest.

\begin{deluxetable}{cc}
\tablecaption{\small Completeness Limits. \label{tab4}}
\tablewidth{0pt}
\tablecolumns{2}
\tablehead{
\colhead{Frame} & \colhead{$50\%$  Limit}}
\startdata
\Ta & $T_1 = 22.6$ \nl
\Tbc & $T_1 = 22.4$ \nl
\Ia & $I = 21.6$ \nl
\Ic & $I = 22.2$ \nl
\Ide & $I = 22.4$ \nl
\If & $I = 22.3$ \nl
\enddata
\end{deluxetable}


For galaxies as distant as NGC 1400 and NGC 1407, individual globular
clusters are indistinguishable from stars under typical seeing conditions.
A galaxy's GCS reveals itself as a statistical overabundance of faint,
star-like features concentrated around the galaxy.
An effective means of identifying non-stellar
objects employs the CLASSIFY routine (Harris 1990).  This routine
calculates radial moments using weighted intensity sums around each object
(see Harris et al. 1991 and references therein).
We chose to use the effective radius parameter $r_{-2}$ as a
discriminator:

\begin{equation}
r_{-2} = \left(\frac{\sum\left[I_j/(r^2 + 0.5)\right]}{\sum I_j}\right)^{-1/2},
\end{equation}

\noindent where $r$ is the radial distance of the $j^{\rm th}$ pixel from
the object centroid, and $I_j$ is the intensity of the $j^{\rm th}$ pixel after
subtraction of the the local sky value.  The summations were performed
over those pixels within a maximum radius of 10 pixels which had
intensities greater than the $3.5\sigma$ detection threshold.
For objects of a given magnitude, non-stellar objects such as
faint background galaxies tend to exhibit more
extended wings and faint cores, thus deviating towards larger values
of $r_{-2}$.
Upper limits were chosen such that roughly $95\%$ of the artificial stars
were correctly classified.
In total, between $20\%$ and $40\%$ of the detected objects in each
frame were identified as non-stellar and were removed from each
photometry list.


Photometric calibrations were performed in the standard fashion.
Fitting of transformation equations to the photometric magnitudes
from short exposures of standard star fields (Geisler 1996,
Porter\footnotemark\footnotetext{Unpublished:  IRAF PHOTCAL standards
database.}) observed during the same nights as the program fields yielded:

\begin{displaymath}
t_{1(ap)} - T_1 = 1.176 + 0.090\;(C-T_1) + 0.048\;X \\
\end{displaymath}
\begin{displaymath}
i_{(ap)} - I = 0.268 + 0.025\;(V-I) + 0.114\;X,
\end{displaymath}

\noindent where $t_{1(ap)}$ and $i_{(ap)}$ are instrumental magnitudes,
and $T_1$ and $I$ are the known magnitudes of the standards.  Since
only $T_1$ and $I$ frames were obtained
for the program data, the colour indices $C-T_1$ and $V-I$ were set to
1.5 and 1.0 respectively, typical values for globular clusters.  The rms
residuals for the transformations are $\pm0.05$ in $T_1$ and $\pm0.04$ in $I$.

\section{Globular Cluster Luminosity Functions}
\label{sec:gclf}

\subsection{Determining the GCLFs}

The observed globular cluster luminosity functions for NGC 1400 and
NGC 1407 were obtained by binning the objects
in the final photometry list into 0.4 mag intervals.  We subtract
the local sky LF (containing field objects and none of the GCS) from
the observed LF (comprising both GCs and field objects) to reveal the GCLF
for the target galaxy.  No separate background
frames were observed, hence the background density of objects was estimated
from outer regions of the target galaxy frames where
the GC density has dropped to zero (as determined from the radial profiles
presented in the next section).

The galaxy LF and that for the background region were
divided by their respective completeness corrections in order to compensate
for increasing detection incompleteness at fainter magnitudes.
Subtraction of the corrected background LF from the galaxy LF yields
the globular cluster LF for each frame.  The error bars on the
luminosity distributions
reflect the uncertainty on the inferred number of objects per magnitude bin:

\begin{equation}
\sigma_n^2 \approx \left[ \frac{n_{\rm obs}}{f^2} + \frac{(1-f)n^2_{\rm obs}}
{n_{\rm add}f^3}\right],
\label{eq:lfunc}
\end{equation}

\noindent(Bolte 1989).  Since the completeness ($f$) is the ratio of
the observed number of objects ($n_{\rm obs}$) to the number inferred
in a magnitude bin ($n$), Eq.~\ref{eq:cfunc} allows us to
rearrange Eq.~\ref{eq:lfunc} in terms of the known quantities for each bin
to obtain:

\begin{equation}
\sigma_n^2\approx \frac{n}{f}\left(1+ \sigma_f^2\frac{n}
{f}\right).
\end{equation}

Once the final cluster luminosity distributions of each data frame
were calculated, the observed GCLFs corresponding to the same galaxy
and filter were averaged, weighted by the uncertainty on the number
counts.  Unfortunately, for NGC 1400, the CTIO $T_1$ frames in best
seeing were not long exposures (see Table~\ref{tab3}), while the
deeper exposures were in poorer seeing than enjoyed at CFHT.  For
these reasons the NGC 1400 GCLF was not well delineated in $T_1$; we
chose to work with the $I$ observations exclusively.  The final
background-subtracted, completeness-corrected and averaged GCLFs in
$T_1$ and $I$ for NGC 1407 and in $I$ for NGC 1400 are presented in
Figures~\ref{fig2}, \ref{fig3} and \ref{fig4}.

\begin{figure}[ht]
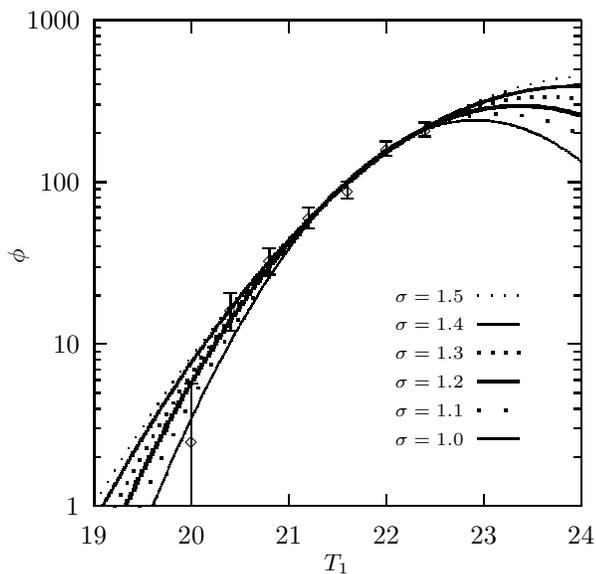

\setlength{\unitlength}{0.240900pt}
\ifx\plotpoint\undefined\newsavebox{\plotpoint}\fi
\sbox{\plotpoint}{\rule[-0.200pt]{0.400pt}{0.400pt}}%
\hspace{-0.3in}

\caption{\small The GCLF of NGC 1407 in $T_1$ with Gaussian fits
for a range of dispersions.}
\label{fig2}
\end{figure}

\begin{figure}[ht]
\setlength{\unitlength}{0.240900pt}
\ifx\plotpoint\undefined\newsavebox{\plotpoint}\fi
\sbox{\plotpoint}{\rule[-0.200pt]{0.400pt}{0.400pt}}%
\hspace{-0.3in}%
\caption{\small The GCLF of NGC 1407 in $I$ with Gaussian fits
for a range of dispersions.}
\label{fig3}
\end{figure}

\begin{figure}[ht]
\setlength{\unitlength}{0.240900pt}
\ifx\plotpoint\undefined\newsavebox{\plotpoint}\fi
\hspace{-0.3in}%
\caption{\small The GCLF of NGC 1400 in $I$ with Gaussian fits
for a range of dispersions.}
\label{fig4}
\end{figure}

The globular cluster luminosity function is defined as the number
of clusters per unit magnitude ($m$) and can be described to first order by a
Gaussian function:

\begin{equation}
\phi(m) = A\exp{ \left(\frac{-(m-m^0)^2}{2\sigma^2}\right) },
\label{eq:gclf}
\end{equation}

\noindent where $A$ is a normalization factor, $m^0$ represents the
turnover (peak) magnitude and $\sigma$ is the dispersion (Hanes 1977).

To fit the Gaussian function of Eq.~\ref{eq:gclf}, a nonlinear least-squares
fitting procedure was applied.  Since the data does not extend past the
turnover magnitude, it was not possible to leave all three parameters
($A$, $m^0$, $\sigma$) unconstrained during the fitting.
Consequently, a series of values of $1.0\lesssim\sigma\lesssim 1.5$
were adopted and held fixed while the scale factor and turnover magnitude
were permitted to vary.  These values of $\sigma$ span the range
typically found for GCLFs (Harris 1991).
The best-fit results are shown in Tables~\ref{tab5}, \ref{tab6} and \ref{tab7}.
Note that the uncertainties quoted here reflect the formal
errors associated with the least-squares fit.

\begin{deluxetable}{cccc}
\tablewidth{0pt}
\tablecolumns{4}
\tablecaption{\small GCLF fit parameters for NGC~1407 in $T_1$. \label{tab5}}
\tablehead{
\colhead{$\sigma$} & \colhead{A} & \colhead{$T^0_1$} &
\colhead{$\chi^2$}
}
\startdata
1.0 & $241 \pm 15$ & $22.92 \pm 0.07$ & 5.2 \nl
1.1 & $264 \pm 17$ & $23.13 \pm 0.07$ & 3.4 \nl
1.2 & $296 \pm 22$ & $23.37 \pm 0.08$ & 2.8 \nl
1.3 & $336 \pm 29$ & $23.64 \pm 0.09$ & 3.0 \nl
1.4 & $389 \pm 41$ & $23.93 \pm 0.11$ & 3.6 \nl
1.5 & $457 \pm 59$ & $24.24 \pm 0.13$ & 4.5 \nl
\enddata
\end{deluxetable}

\begin{deluxetable}{cccc}
\tablewidth{0pt}
\tablecolumns{4}
\tablecaption{\small GCLF fit parameters for NGC~1407 in $I$. \label{tab6}}
\tablehead{
\colhead{$\sigma$} & \colhead{A} & \colhead{$I^0$} & \colhead{$\chi^2$}
}
\startdata
1.0 & $313 \pm 33$ & $22.72 \pm 0.11$ & 6.3 \nl
1.1 & $358 \pm 45$ & $22.97 \pm 0.12$ & 4.9 \nl
1.2 & $418 \pm 64$ & $23.25 \pm 0.15$ & 4.3 \nl
1.3 & $498 \pm 92$ & $23.59 \pm 0.17$ & 4.3 \nl
1.4 & $615 \pm 133$ & $23.89 \pm 0.20$ & 4.5 \nl
1.5 & $749 \pm 194$ & $24.24 \pm 0.23$ & 5.1 \nl
\enddata
\end{deluxetable}

\begin{deluxetable}{cccc}
\tablewidth{0pt}
\tablecolumns{4}
\tablecaption{\small GCLF fit parameters for NGC~1400 in $I$. \label{tab7}}
\tablehead{
\colhead{$\sigma$} & \colhead{A} & \colhead{$I^0$} & \colhead{$\chi^2$}
}
\startdata
1.0 & $111 \pm 25$  & $23.22 \pm 0.21$ & 15.4 \nl
1.1 & $130 \pm 37$  & $23.50 \pm 0.25$ & 15.6 \nl
1.2 & $157 \pm 53$  & $23.80 \pm 0.30$ & 15.9 \nl
1.3 & $194 \pm 79$  & $24.13 \pm 0.35$ & 16.1 \nl
1.4 & $244 \pm 118$ & $24.49 \pm 0.41$ & 16.4 \nl
1.5 & $316 \pm 178$ & $24.88 \pm 0.47$ & 16.7 \nl
\enddata
\end{deluxetable}

\subsection{Distance Determinations}

Comparing the observed brightness distributions for the globular
cluster systems of NGC~1400 and NGC~1407 with those known from local
calibrators, we can calculate distances to these galaxies.  With an
adopted absolute turnover magnitude of $M^0_{\mbox{\tiny V}} = -7.27
\pm 0.23$ (a mean value for the 10 largest galaxies in Table~2 of
Harris 1991) and assuming $\sigma=1.2$ for the GCLF shapes, we
estimate distances of $17.6 \pm 3.1$~Mpc to NGC~1407 and $25.4 \pm
7.0$~Mpc to NGC~1400 (Table~\ref{tab8}).  Note that the turnover magnitudes
have been corrected for foreground absorption.  For NGC~1407,
we adopted galactic extinction corrections of A$_{T_{1}}=0.11$ and A$_I=0.06$,
based on a value of A$_B=0.17$ (RC3).  NGC~1400's value of A$_B=0.14$ 
yielded corrections of A$_{T_{1}}=0.09$ and A$_I=0.06$ for this galaxy.
The distances we have calculated in Table~\ref{tab8}
are inconsistent with the hypothesis that NGC~1400 is a foreground
galaxy roughly 3 times closer than NGC~1407, as implied by its
anomalously low recessional velocity.

\begin{deluxetable}{crccc}
\tablewidth{405pt}
\tablecolumns{5}
\tablecaption{\small Galaxy distances assuming $\sigma=1.2$, corrected for
absorption. \label{tab8}}
\tablehead{
\colhead{Galaxy} & \colhead{$m^\circ$} & \colhead{$V^\circ$} &
\colhead{$(V^0 - M^0)$} & \colhead{$d$ (Mpc)}
}
\startdata
NGC~1407 & $T_1^0=23.26 \pm 0.08$ & $23.71 \pm 0.08$
   & $30.98 \pm 0.31$ & $15.7\;(^{+2.4}_{-2.1})$ \nl
~ & $I^0=23.19 \pm 0.15$ & $24.19 \pm 0.15$
   & $31.46 \pm 0.38$ & $19.5\;(^{+3.7}_{-3.1})$ \nl
NGC~1400 & $I^0=23.75 \pm 0.30$ & $24.75 \pm 0.30$
   & $32.02 \pm 0.53$ & $25.4\;(^{+7.0}_{-5.5})$ \nl
\enddata
\end{deluxetable}

Further experimentation shows that the choice of dispersion ($\sigma$)
has little bearing on the relative distances obtained for these
galaxies.  The distance ratios fall within the range
$d(N1400)/d(N1407)=1.31 \pm 0.04$ regardless of adopted $\sigma$,
suggesting that NGC~1400 is in the background with respect to
NGC~1407.  Adopting the worst-case scenario in which $\sigma=1.5$ for
NGC~1407 and $\sigma=1.0$ for NGC~1400, we find that
$d(N1400)/d(N1407)=0.63$, still placing NGC~1400 considerably further
than the distance implied by its low recessional velocity.  From this
we may again conclude that, within the reasonable range of $\sigma$,
our observations of the GCS of NGC~1400 are not consistent with it
being a factor of 3 closer than NGC~1407.

If NGC~1400 were at a distance of $5.49h^{-1}$~Mpc, the turnover
magnitude of the globular cluster luminosity function would be
expected to appear at $T_1^0=21.6$ and $I^0=21.0$, assuming $h=0.8$.
With NGC~1400 GCS luminosity distribution data down to completeness
limits of $T_1=22.6$ and $I=22.4$, our GCLFs should thus extend well
beyond the turnover in such a case.  Since the observed GCLFs for this
galaxy do not show any evidence of having reached the turnover
magnitude, this reaffirms our conclusion that NGC~1400 is not a
foreground galaxy one third the distance of Eridanus~A.

\section{Cluster Radial Profiles}
\label{sec:rp}

The surface density distribution of GCs plotted as a function of galactocentric
radius reveals the projection of the spatial structure of the GCS.
The annular bins used to construct the profiles were 75 pixels wide
and concentric on the galaxy centroids.
The number counts in each radial bin, corrected for completeness
and divided by the total area of the annulus sampled,
were taken to represent the surface density at the geometric mean annular
radius ($r=\sqrt{r_{\rm in}r_{\rm out}}$).
Subtracting the local background density ($\sigma_{\rm bgd}$),
found by adopting the mean value of the surface density beyond the radius
at which the radial profile is no longer decaying, we obtain the
distribution of the GC population alone.

Non-linear least-squares fitting of the following relations were
performed on the radial profiles:

\begin{mathletters}
\begin{equation}
\log{\sigma_{\rm GC}} = a + b\;r^{1/4}
\label{eq:deV}
\end{equation}
\begin{equation}
\log{\sigma_{\rm GC}} = c + d\log{r}.
\label{eq:powerlaw}
\end{equation}
\end{mathletters}

\noindent The first model profile is a form of the empirical de~Vaucouleurs
law, and the second represents a scale-free power law (Harris 1986).
The curves superimposed upon the radial profiles shown in
Figures~\ref{fig5} to \ref{fig8} represent the
best fits to the data with the parameters
and formal errors listed in Table~\ref{tab9}.
In all of the radial profile plots, we can clearly identify the presence of
a centrally concentrated globular cluster system.

Surface intensity profiles of the galaxy halo light are also provided
on the GC radial profile plots.  These halo profiles were determined
by fitting elliptical isophotes over the galaxy region, yielding the
relative intensity of light as a function of radius.  After
subtracting the local sky level, the halo intensity was arbitrarily
scaled to the amplitude of the GCS radial distribution for easy
comparison.

\begin{figure}[ht]
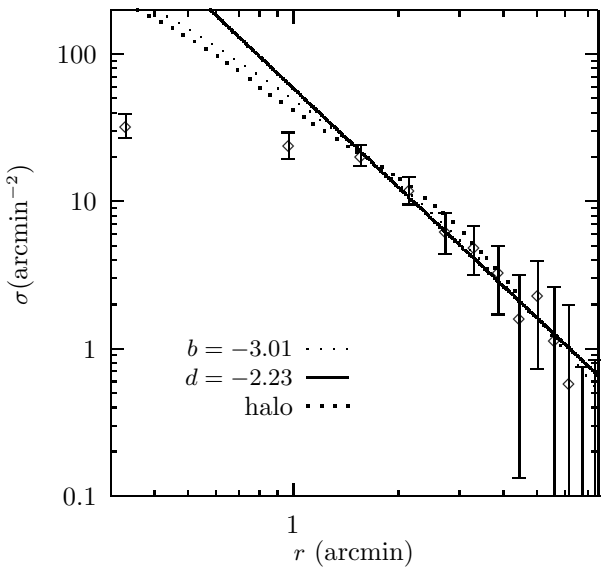

\setlength{\unitlength}{0.240900pt}
\ifx\plotpoint\undefined\newsavebox{\plotpoint}\fi
\sbox{\plotpoint}{\rule[-0.200pt]{0.400pt}{0.400pt}}%
\hspace{-0.3in}

\caption{\small The background-subtracted radial density
profile of the NGC~1407 GCS (in $T_1$) with de~Vaucouleurs and
scale-free power law fits provided (slopes given by $b$ and $d$
respectively).  Also shown is the galaxy halo intensity profile.  The
background density level was found to be $\sigma_{\rm bgd} = 5.9 \pm 0.3$.}
\label{fig5}
\end{figure}

\begin{figure}[ht]
\setlength{\unitlength}{0.240900pt}
\ifx\plotpoint\undefined\newsavebox{\plotpoint}\fi
\sbox{\plotpoint}{\rule[-0.200pt]{0.400pt}{0.400pt}}%
\hspace{-0.3in}

\caption{\small The NGC~1407 GCS radial density profile (in $I$), as in
Figure~\protect\ref{fig5}.  The background density was $\sigma_{\rm
bgd} = 9.8 \pm 0.6$.}
\label{fig6}
\end{figure}

\begin{figure}[ht]
\setlength{\unitlength}{0.240900pt}
\ifx\plotpoint\undefined\newsavebox{\plotpoint}\fi
\sbox{\plotpoint}{\rule[-0.200pt]{0.400pt}{0.400pt}}%
\hspace{-0.3in}%
\caption{\small The NGC~1400 GCS radial density profile (in $T_1$), 
as in Figure~\protect\ref{fig5}.  The background density was $\sigma_{\rm bgd} = 5.7 \pm 0.4$.}
\label{fig7}
\end{figure}

\begin{figure}[ht]
\setlength{\unitlength}{0.240900pt}
\ifx\plotpoint\undefined\newsavebox{\plotpoint}\fi
\sbox{\plotpoint}{\rule[-0.200pt]{0.400pt}{0.400pt}}%
\hspace{-0.3in}%
\caption{\small The NGC~1400 GCS radial density profile (in
$I$), as in Figure~\protect\ref{fig5}.  The background density was
$\sigma_{\rm bgd} = 10.4 \pm 0.7$.}
\label{fig8}
\end{figure}

\begin{deluxetable}{cccccc}
\tablewidth{0pt}
\tablecaption{\small GC radial profile best-fit parameters.  \label{tab9}}
\tablecolumns{6}
\tablehead{
\colhead{~} & \colhead{~} &
\multicolumn{2}{c}{$\log{\sigma_{\rm GC}} = a + b\; r^{1/4}$} &
\multicolumn{2}{c}{$\log{\sigma_{\rm GC}} = c + d \log{r}$} \nl
\colhead{Galaxy} & \colhead{Filter} & \colhead{$a$} & \colhead{$b$} &
\colhead{$c$} & \colhead{$d$}
}
\startdata
NGC 1407 & $T_1$ & $4.70 \pm 0.56$ & $-3.01 \pm 0.47$ & $1.77 \pm 0.12$
 & $-2.23 \pm 0.34$ \nl
~ & $I$ & $4.33 \pm 0.63$ & $-2.83 \pm 0.63$ & $1.46 \pm 0.11$
 & $-1.74 \pm 0.38$ \nl
NGC 1400 & $T_1$ & $4.62 \pm 1.27$ & $-3.71 \pm 1.22$ & $0.90 \pm 0.12$
 & $-2.35 \pm 0.78$ \nl
~ & $I$ & $3.79 \pm 0.41$ & $-2.81 \pm 0.46$ & $0.96 \pm 0.09$
 & $-1.57 \pm 0.34$ \nl
\enddata
\end{deluxetable}

\section{Specific Frequencies}
\label{sec:sp}

The specific frequency ($S_N$) of a galaxy is defined as the number
of globular clusters per unit halo light, normalized to an absolute
magnitude of $M_V= 15$:
\begin{equation}
S_N\equiv N_t \times 10^{0.4(M_V + 15)}
\label{eq:sn}
\end{equation}

\noindent where the total population of GCs surrounding a galaxy
is given by $N_t$ (Harris \& van den Bergh 1981).  This parameter
conveniently reflects the size of the GC population, independent
of galaxy luminosity.

To calculate specific frequency we must first estimate
the total number of clusters which make up the GCS of the galaxy.
We use the radial density profiles from
\S\ref{sec:rp} to count the total number of the clusters
surrounding the galaxy down to the limiting magnitude ($N_{obs}$).
We divide this result by the fraction of the GCLF which was
observed to infer the total number of clusters in the system.
To define the shapes of the GCLFs, we adopt a
Gaussian form (Eq.~\ref{eq:gclf}) assuming $\sigma=1.2$ and
the best-fit parameters from Tables~\ref{tab5}, \ref{tab6} and \ref{tab7}.

Absolute $V$ magnitudes are also required in order to determine $S_N$.
According to RC3, NGC 1407 has a corrected total blue magnitude
$B_{\mbox{\tiny T}} = 10.5 \pm 0.2$ and colour $(B-V)_{\mbox{\tiny T}}
= 0.97 \pm 0.01$, yielding $V_{\mbox{\tiny T}} = 9.5 \pm 0.2$.  For
NGC 1400, the RC3 values are $B_{\mbox{\tiny T}} = 11.87 \pm 0.13$ and
$(B-V)_{\mbox{\tiny T}} = 0.92 \pm 0.01$, giving $V_{\mbox{\tiny T}} =
10.95 \pm 0.13$ for this galaxy.  To translate $V_{\mbox{\tiny T}}$
into absolute magnitude $M_{\mbox{\tiny V}}$ we must adopt distances
to the galaxies in question.  For both galaxies, we assume a distance
of $20.5 \pm 1.2$~Mpc based on the mean recessional velocity of the
Eridanus~A galaxies and H$_0= 80$~km~s$^{-1}$~Mpc$^{-1}$.  The results
of the specific frequency calculations appear in Table~\ref{tab10}.
The specific frequencies calculated for NGC~1407 in the two filters
are in agreement within their respective uncertainties, and we
therefore adopt a mean value of $S_N = 4.0 \pm 1.3$ for NGC~1407 along
with that of $S_N=5.2 \pm 2.0$ for NGC~1400.

\begin{deluxetable}{cccccc}
\tablewidth{0pt}
\tablecolumns{6}
\tablecaption{\small Specific Frequencies \label{tab10}}
\tablehead{
\colhead{Galaxy} & \colhead{Filter} & \colhead{$N_{\rm obs}$} &
\colhead{$N_t$} & \colhead{$M_V$} & \colhead{$S_N$}
}
\startdata
NGC~1407 & $T_1$ & $629 \pm 76$ & $2296 \pm 279$ & $-22.06 \pm 0.33$
 & $3.4 \pm 1.1$ \nl
~ & $I$ & $556 \pm 61$ & $2985 \pm 327$ & $-22.06 \pm 0.33$ & $4.5 \pm 1.4$ \nl
NGC~1400 & $I$ & $106 \pm 32$ & $922 \pm 280$ & $-20.61 \pm 0.26$
 & $5.2 \pm 2.0$ \nl
\enddata
\end{deluxetable}

\section{Discussion}

\subsection{The Question of Distance}

The primary motivation of this study was to determine whether NGC~1400 is
at a distance comparable to that of NGC~1407 and the rest of Eridanus~A,
or if it is instead a foreground galaxy as implied by its anomalously low
recessional velocity.
The globular cluster luminosity functions are not consistent with
the foreground placement of NGC~1400, which indicates that this galaxy
has a large peculiar velocity with respect to Eridanus~A, the origin of
which remains unknown.

We may now add globular cluster systems to the evidence
that NGC~1400 does indeed lie at roughly the distance of Eridanus~A,
and should be counted as a member of the cluster.
The distances we have calculated assuming $\sigma=1.2$ for the GCLF
are slightly higher but still within two standard deviations from those derived
by Tonry (1991) of $16.3 \pm 1.0$~Mpc (NGC~1400) and
$16.4 \pm 1.0$~Mpc (NGC~1407), and are well within the range of potential
distances previously quoted in the literature (see Table~\ref{tab2}).

\subsection{GCS Shapes and Specific Frequencies}

The spatial profiles of the globular cluster systems of NGC~1400 and
NGC~1407 shown in \S\ref{sec:rp} seem rather unexceptional in shape.
The radial distributions in Figures~\ref{fig5} and \ref{fig6} reveal
that the projected density of NGC~1407's GCS falls off with the same
slope as the halo light intensity for this central elliptical galaxy
of Eridanus~A.  Figures~\ref{fig7} and \ref{fig8} indicate that the
light profile for the NGC~1400 halo may be slightly steeper than the
cluster density distribution, but this effect is not uncommon.  It is
somewhat unusual for GCS radial profiles to follow the halo light
profile as closely as seen for NGC~1407; generally, the cluster
surface density is more distended (Harris 1991).  According to Merritt
(1983, 1984), the cluster population and halo structure of a system
dominated by dark matter will remain largely unchanged (to first
order), not significantly altered by galaxy interaction processes.  We
are therefore likely seeing the cluster distributions as they were
originally formed, with GCSs which are roughly as centrally
concentrated as the halo stars.

There exists an apparent relationship between the integrated $V$
magnitude of a galaxy and the shape of its GCS density profile (Harris
1986).  The parameter generally used to describe the shape of the
radial distribution is the slope of the logarithmic profile.  This
parameter was given the label ``$d$'' in the power-law fitting of
Section~\ref{sec:rp}, but is commonly referred to as $\alpha$.  A plot
of the slope parameter as a function of galaxy magnitude is shown in
Figure~\ref{fig9}, with data and references presented in Table~\ref{tab11}.  A
linear fit to the data yields the relation:
\begin{mathletters}
\begin{equation}
\alpha=-7.93-(0.29 \pm 0.03)M^{\mbox{\tiny T}}_{\mbox{\tiny V}}
\label{eq:shape}
\end{equation}
\noindent This is consistent with the relation as determined in Harris 1993
(see correction appearing in Kaisler et al. 1996).

\begin{figure}[ht]
\hspace*{\fill}\psfig{file=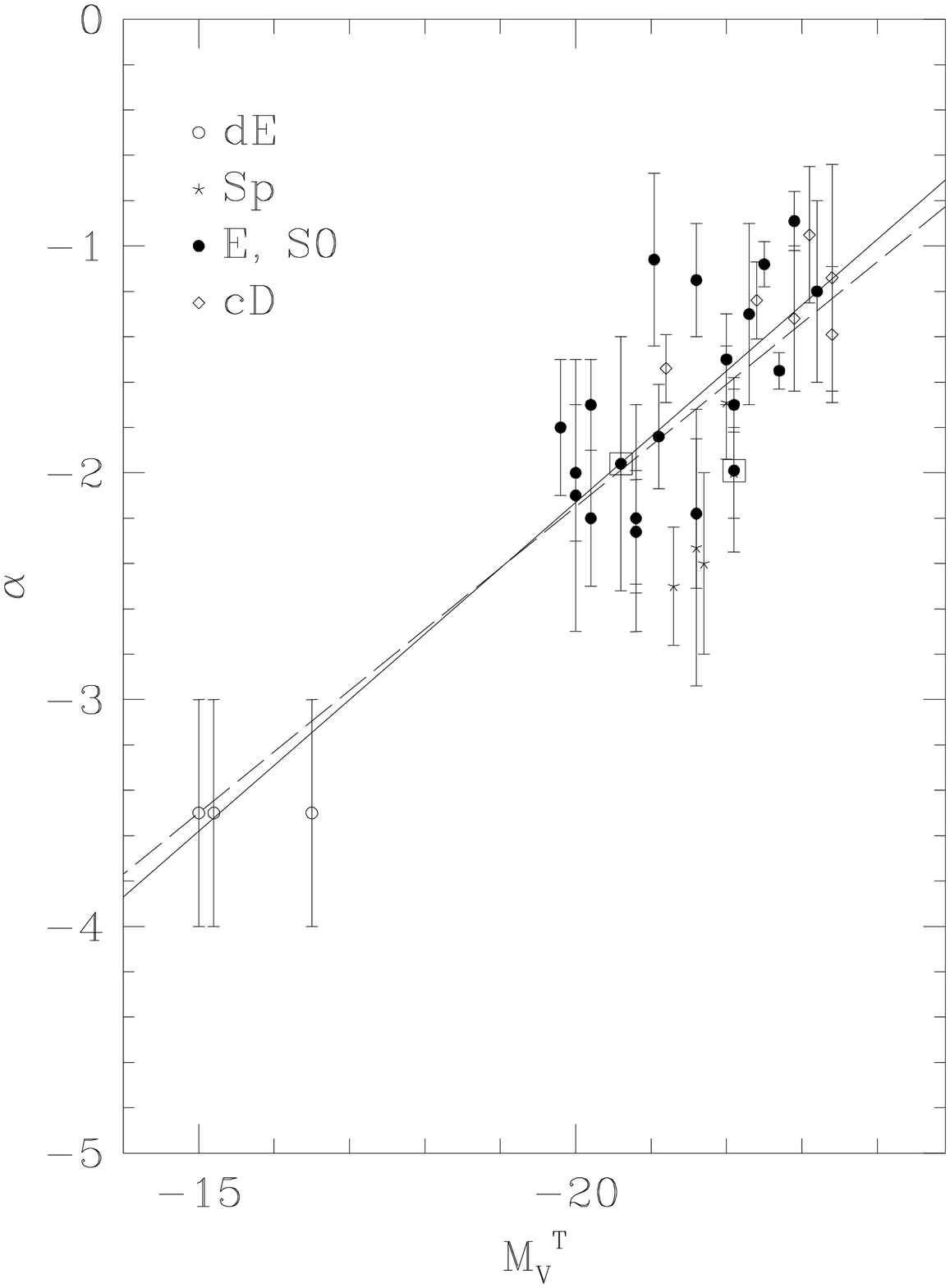,height=3in,width=3in}
\hspace*{\fill}\caption{\small The relationship between slope of the GCS
radial distribution and total galaxy V magnitude.  The solid line
represents a linear fit to the data including all data points
(Eq.~\protect\ref{eq:shape}), and the dashed line indicates the fit if
one does not include dwarf or spiral galaxies in the sample
(Eq.~\protect\ref{eq:shapeb}).  The data points enclosed by large open
squares highlight the values obtained for NGC~1400 and NGC~1407 in the
present study.}
\label{fig9}
\end{figure}

\begin{deluxetable}{ccccc}
\tablewidth{0pt}
\tablecaption{\small Correlation between radial density and galaxy luminosity.
\label{tab11}}
\tablecolumns{5}
\tablehead{
\colhead{Galaxy} & \colhead{Type} &
\colhead{$M^{\mbox{\tiny T}}_{\mbox{\tiny V}}$} & \colhead{$\alpha$} &
\colhead{Source}
}
\startdata
Milky Way & Sbc & -21.3  & $-2.5 \pm 0.26$  & {H81, H76} \nl
NGC~147 & dE    & -15.0  & $-3.5 \pm 0.5$   & {H86, HR79} \nl
NGC~185 & dE    & -15.2  & $-3.5 \pm 0.5$   & {H86, HR79} \nl
NGC~205 & dE    & -16.5  & $-3.5 \pm 0.5$   & {H86, HR79} \nl
NGC~224 & Sb    & -21.7  & $-2.4 \pm 0.4$   & {H81, HR79} \nl
NGC~524 & E/S0  & -22.1  & $-1.7 \pm 0.12$  & {HH85} \nl
NGC~1052 & E    & -20.8  & $-2.26 \pm 0.27$ & {HH85} \nl
NGC~1374 & E    & -19.8  & $-1.8 \pm 0.3$   & {KKHRIQ96} \nl
NGC~1379 & E    & -20.0  & $-2.1 \pm 0.6$   & {HH86b, KKHRIQ96} \nl
NGC~1387 & S0   & -20.2  & $-2.2 \pm 0.3$   & {HH86b, KKHRIQ96} \nl
NGC~1399 & E    & -21.2  & $-1.54 \pm 0.15$ & {BHH91, KKHRIQ96} \nl
NGC~1400 & E/S0 & -20.6  & $-1.96 \pm 0.56$ & {This Paper} \nl
NGC~1404 & E    & -20.8  & $-2.2 \pm 0.5$   & {HH86b} \nl
NGC~1407 & E    & -22.1  & $-1.99 \pm 0.36$ & {This Paper} \nl
NGC~1427 & E    & -20.0  & $-2.0 \pm 0.3$   & {KKHRIQ96} \nl
NGC~3115 & S0   & -21.1  & $-1.84 \pm 0.23$ & {HH86a} \nl
NGC~3311 & cD   & -22.4  & $-1.24 \pm 0.17$ & {H86} \nl
NGC~3842 & E    & -23.2  & $-1.2 \pm 0.4$   & {BH92} \nl
NGC~4073 & cD   & -23.1  & $-0.95 \pm 0.3$  & {BH94} \nl
NGC~4278 & E    & -20.8  & $-2.26 \pm 0.23$ & {Hv81} \nl
NGC~4365 & E    & -21.6  & $-1.15 \pm 0.25$ & {HAPv91} \nl
NGC~4472 & E    & -22.9  & $-0.89 \pm 0.13$ & {HAPv91} \nl
NGC~4486 & E    & -22.7  & $-1.55 \pm 0.08$ & {H86}  \nl
NGC~4565 & Sb   & -21.6  & $-2.33 \pm 0.61$ & {FHPH95, H81, vH82} \nl
NGC~4594 & Sa   & -22.1  & $-2.0 \pm 0.2$   & {HHH84} \nl
NGC~4649 & E    & -22.5  & $-1.08 \pm 0.10$ & {HAPv91} \nl
NGC~4944 & E    & -21.04 & $-1.06 \pm 0.38$ & {FHPH95} \nl
NGC~5018 & E    & -22.3  & $-1.3 \pm 0.4$   & {HK96} \nl
NGC~5128 & Ep   & -22.0  & $-1.5 \pm 0.2$   & {HHHC84} \nl
NGC~5170 & Sb   & -22.0  & $-1.69 \pm 0.25$ & {FHHB90} \nl
NGC~5481 & E    & -20.2  & $-1.7 \pm 0.2$   & {MR95} \nl
NGC~5813 & E    & -21.6  & $-2.18 \pm 0.33$ & {Hv81} \nl
NGC~7768 & cD   & -22.9  & $-1.32 \pm 0.32$ & {HPM95} \nl
UGC~9799 & cD   & -23.4  & $-1.39 \pm 0.30$ & {HPM95} \nl
UGC~9958 & cD   & -23.4  & $-1.14 \pm 0.50$ & {HPM95} \nl
\enddata
\end{deluxetable}

\sloppy{The two galaxies in this study follow the overall trend whereby
higher-luminosity galaxies exhibit more extended globular cluster
systems.  The measured slopes, $\alpha{\rm (NGC 1407)}=-1.99 \pm 0.36$
and $\alpha{\rm (NGC 1400)}=-1.96 \pm 0.56$, compare reasonably well
with the predicted values of $\alpha=-1.53$ and $-1.95$ respectively.
If we do not include the data points for the dwarf ellipticals or the
spiral galaxies, a linear regression yields}
\begin{equation}
\alpha=-7.55-(0.27 \pm 0.05)M^{\mbox{\tiny T}}_{\mbox{\tiny V}}.
\label{eq:shapeb}
\end{equation}
\end{mathletters}
\noindent The omission does not significantly alter the relation; both
of the linear fits above are shown in Figure~\ref{fig9}.

The specific frequencies calculated for NGC~1407 ($S_N=4.0 \pm 1.3$)
and NGC~1400 ($S_N=5.2 \pm 2.0$) are not unusual.  Normal
elliptical galaxies generally have $S_N\sim 2-5$, with variations from
this range most likely attributable to differences in galaxy
environment and GC formation mechanisms.

\subsection{GC Colours}

Although the $T_1$ and $I$ filters do not provide a sufficient
baseline to obtain cluster metallicities, we can estimate mean colours
for the globulars surrounding NGC~1407 and NGC~1400.  Using the
regions of overlap between the $T_1$ and $I$ images, we determine
$T_1-I$ colours for the GCs detected in both filters.  The colour
distributions are shown in Figure~\ref{fig10}; the lack of a well-defined peak
in the NGC~1400 GCS colour distribution is likely due to the poor
quality of the $T_1$ photometry for this galaxy, with its smaller
cluster population.  NGC~1407's cluster population is found to have a
mean of $T_1-I=0.53$ and a median value of $0.54$, while for NGC~1400
we obtain mean and median colours of $T_1-I=0.54$ and $0.55$,
respectively.  These values are consistent with typical GC colours
observed in other galaxies, a finding which confirms the identification
of the GCSs.

\begin{figure}[ht]
\hspace*{\fill}\psfig{file=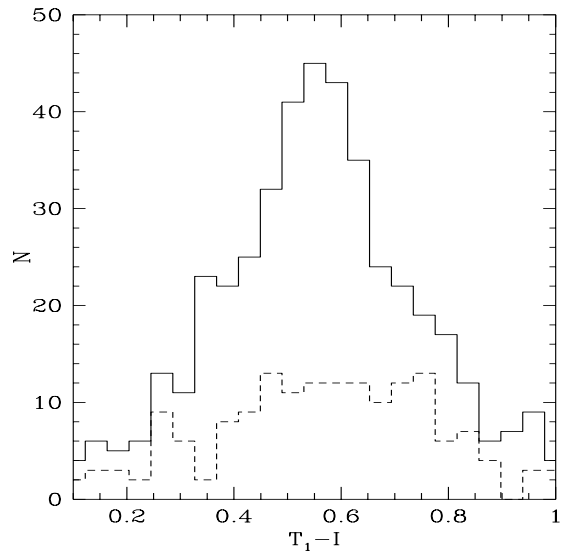,height=3in,width=3in}
\hspace*{\fill}\caption{\small Colour distribution for the globular clusters
surrounding NGC~1407 (solid line) and NGC~1400 (dashed line).}
\label{fig10}
\end{figure}

\section{Conclusions}

The principal results of this study of the globular cluster systems of
NGC~1400 and NGC~1407 can be summed up in the following points:

\begin{enumerate}

\item From the shapes of the globular cluster luminosity functions, we
determine distances to these galaxies which place NGC~1400 at or
beyond the distance of the Eridanus~A group.  This finding is in
agreement with conclusions made using other methods previously cited in
the literature.

\item The shapes of the GCS radial density profiles and the specific
frequencies of the two systems reveal no obvious abnormalities.
This implies that, if Eridanus~A is as dominated by dark matter as its
estimated $M/L$ value indicates, no anomalies are evident from the GC
spatial distributions and population sizes of its two largest
galaxies.

\end{enumerate}

With a distance comparable to that of NGC~1407, NGC~1400 must have a
high peculiar velocity in order to
account for its exceptionally low radial velocity.
Gould (1993) demonstrates that if the distance of NGC~1400 is consistent
with that of Eridanus~A, it must be bound to the sub-cluster due to a lack of
other nearby mass concentrations large enough to generate its high
peculiar motion.  The exact cause for the large peculiar velocity remains
as yet unknown.  It is possible that NGC~1400 has a
large component of its velocity moving it towards the
core of Eridanus~A (ie:  it has a large transverse velocity), reducing
its net radial velocity.  Perhaps the velocity dispersion of the cluster
has been severely underestimated since we only have velocity
data for 10 of the 50 or so members,
and that by some coincidence the other galaxies in the sample have
significant velocity components perpendicular to our line of sight.
If this is the case, NGC~1400's high peculiar velocity might not
be particularly anomalous.

The second enigma surrounding the Eridanus~A sub-cluster is its abnormally
high $M/L$ ratio.  It is possible that it is merely a dark cluster --
many such clusters could exist which have so far avoided detection.
More expansive surveys at higher limiting magnitudes in combination
with reliable cluster-finding algorithms may reveal
the presence of more dark clusters.  The question remains:
if indeed this cluster contains a great deal of dark matter, where did it
come from, and why is NGC~1400 the only member (so far) to show such a high
peculiar velocity?  Furthermore, why is there no evidence of strange
effects on the radial distributions and population sizes of the member galaxy
GCSs, for example, given this unusual environment?

A more extensive analysis of the GCS of the two Eridanus~A galaxies
NGC~1407 and NGC~1400 could be provided by additional deeper
multicolour photometry as well as spectroscopic observations.  This
data may contribute to a better understanding of the nature of the
galaxies and their environment.  There is an obvious lack of redshift
measurements for the majority of the Eridanus~A galaxies (see Table~4
of Ferguson \& Sandage (1990) for the complete membership list).  This
makes it difficult to determine an accurate $M/L$ ratio for the
cluster, as well as to derive any conclusions regarding the dynamical
processes at work within Eridanus~A.  A more complete database of
member galaxy velocities may shed some light on this dark cluster.

\acknowledgements{The authors wish to extend their gratitude to the
kind and helpful staff at CTIO and CFHT.  This work was supported by
the Natural Sciences and Engineering Research Council of Canada in the
form of operating grants (DAH and WEH) and a PGS~A scholarship (KMP).}


}
\end{document}